\documentclass[iop,apj,tighten]{emulateapj}
\usepackage{amstext}
\usepackage{apjfonts}
\usepackage{amsmath}
\usepackage{amssymb}
\usepackage[breaklinks,colorlinks,citecolor=blue,linkcolor=magenta]{hyperref} 
\usepackage{comment}
\usepackage{mathrsfs}
\usepackage[all]{hypcap} %Links go to figures. This breaks deluxetables; use \capstartfalse \capstarttrue around deluxetables to fix it.
\usepackage{bm}

\shorttitle{Eccentric Black Hole Mergers}
\shortauthors{Samsing, Askar \& Giersz}

\begin{document}

\title{MOCCA-SURVEY Database I: Eccentric Black Hole Mergers During Binary-Single Interactions In Globular Clusters} 
\author{Johan Samsing$^{1}$, Abbas Askar$^{2}$, and Mirek Giersz$^{2}$} 
\altaffiltext{1}{Department of Astrophysical Sciences, Princeton University, Peyton Hall, 4 Ivy Lane, Princeton, NJ 08544, USA}
\altaffiltext{2}{Nicolaus Copernicus Astronomical Center, Polish Academy of Sciences, ul. Bartycka 18, 00-716 Warsaw, Poland}

%%%%%%%%%%%%%%%%%%%%%%%%%%%%%%%%%%%%%%%%%%%%%%%%%%%%%%%%%%%%%%%%%%%%%%%%   
\begin{abstract} 
%%%%%%%%%%%%%%%%%%%%%%%%%%%%%%%%%%%%%%%%%%%%%%%%%%%%%%%%%%%%%%%%%%%%%%%%

We estimate the population of eccentric gravitational wave (GW) binary black hole (BBH) mergers forming
during binary-single interactions in globular clusters (GCs), using $\sim 800$ GC models that were evolved using the MOCCA code for star cluster
simulations as part of the MOCCA-Survey Database I project. 
By re-simulating binary-single interactions (only involving 3 BHs) extracted from this set of GC models using
an $N$-body code that includes GW emission at the 2.5 post-Newtonian level, we find that $\sim 10\%$ of all
the BBHs assembled in our GC models that merge at present time form during chaotic binary-single interactions,
and that about half of this sample have an eccentricity $>0.1$ at $10$ Hz.
We explicitly show that this derived rate of eccentric mergers is $\sim 100$ times higher than one would find with a purely Newtonian $N$-body code.
Furthermore, we demonstrate that the eccentric fraction can be accurately estimated using a simple analytical formalism
when the interacting BHs are of similar mass; a result that serves as the first successful analytical description of eccentric GW mergers
forming during three-body interactions in realistic GCs.

\end{abstract}

%%%%%%%%%%%%%%%%%%%%%%%%%%%%%%%%%%%%%%%%%%%%%%%%%%%%%%%%%%%%%%%%%%%%%%%%

\section{Introduction}

Gravitational waves (GWs) from binary black hole (BBH) mergers have recently
been observed \citep{2016PhRvL.116f1102A, 2016PhRvL.116x1103A, 2016PhRvX...6d1015A, 2017PhRvL.118v1101A,
2017PhRvL.119n1101A}, but how the BBHs formed and merged
is still an open question. Several merger scenarios have been proposed, from
isolated field mergers \citep{2015ApJ...806..263D, 2016ApJ...819..108B,
2016Natur.534..512B} and dynamically assembled cluster mergers \citep{2000ApJ...528L..17P, 2010MNRAS.402..371B, 2013MNRAS.435.1358T,
2014MNRAS.440.2714B, 2016PhRvD..93h4029R,2016MNRAS.458.3075A, 2016ApJ...824L...8R, 
2017MNRAS.464L..36A, 2017MNRAS.469.4665P}, to primordial BH capture mergers \citep{2016PhRvL.116t1301B, 2016PhRvD..94h4013C,
2016PhRvL.117f1101S, 2016PhRvD..94h3504C} and mergers forming
in active galactic nuclei discs \citep{2017ApJ...835..165B, 2017MNRAS.464..946S, 2017arXiv170207818M}, however,
how to observationally distinguish these channels from each other is a major challenge.

One of the promising parameters that both can be extracted from the observed GW waveform, and also seems to differ between different
merger channels, is the BBH orbital eccentricity at a given gravitational wave frequency \citep[e.g.][]{2017arXiv170603776S}. Generally, one finds that dynamically assembled
BBH mergers have a non-negligible probability to appear eccentric at observation, including hierarchical three-body systems
\citep{2017ApJ...836...39S, 2017ApJ...841...77A}, strong binary-single interactions \citep{2014ApJ...784...71S, 2017ApJ...840L..14S,
2017arXiv170603776S, 2017arXiv171107452S, 2017arXiv171204937R} and single-single
interactions \citep{2009MNRAS.395.2127O, 2012PhRvD..85l3005K, 2016PhRvD..94h4013C, 2017arXiv171109989G},
whereas all isolated field mergers are expected to be circular due to late time orbital circularization through GW emission \citep{Peters:1964bc}.

The importance of including general relativity (GR) in the equation-of-motion (EOM) for probing the population of \textit{eccentric BBH mergers} forming in
globular cluster (GCs), was first pointed out by \cite{2017ApJ...840L..14S, 2017arXiv171107452S}, who derived that the rate of eccentric BBH
mergers ($> 0.1$ at $10$ Hz) forming through binary-single interactions
is about $\sim 100$ times higher when GR is included in the EOM, compared to using a purely Newtonian solver. By integrating over the dynamical history of
a typical BBH \cite{2017arXiv171107452S} showed that this implies that at present time up to $\sim 5\%$ of all
BBH mergers will have an eccentricity $> 0.1$ at $10$ Hz. As described in \cite{2006ApJ...640..156G, 2014ApJ...784...71S}, such eccentric mergers
form through two-body GW captures during three-body interactions.

In this paper, we estimate the fraction of eccentric BBH mergers forming through two-body GW captures during binary-single interactions
in GCs, using the data from `MOCCA-SURVEY Database I', which consists of nearly $2000$ GC models dynamically evolved
by the state-of-the-art Monte-Carlo (MC) code \texttt{MOCCA} \citep{Hypki2013,Giersz2013}.
Originally, all the binary-single interactions evolved for
these GC models were performed with the Newtonian code \texttt{fewbody} \citep{Fregeau2004}; however, for this paper we re-simulate these interactions
using a few-body code that includes orbital energy and angular momentum dissipation through GW emission at the 2.5 post-Newtonian (PN)
level \citep{2017ApJ...846...36S}, with the goal of resolving the eccentric fraction.
We further show how the rate of eccentric BBH mergers can be accurately estimated using a simple
analytical formalism recently presented in \cite{2017arXiv171107452S}, which provides valuable insight into the analytical description and understanding of the
relativistic few-body problem. Finally, we note that a similar study by \cite{2017arXiv171204937R} has been done in parallel to our work, but with a completely
different code and dataset. This study finds, as well as we do, excellent agreement with the analytical predictions made by \cite{2017ApJ...840L..14S, 2017arXiv171107452S}.

In Section \ref{sec:Numerical Codes and Data Models} we introduce the MOCCA code and the extensive GC dataset used for this study; `MOCCA-Survey Database I'.
In Section \ref{sec:Numerical and Analytical Methods} we describe our numerical and analytical approaches for estimating the
fraction of eccentric BBH mergers forming in `MOCCA-Survey Database I'. Results are given in Section \ref{sec:Results}, and conclusions
in Section \ref{sec:Conclusion}.

\section{Codes and Data Models}\label{sec:Numerical Codes and Data Models}

In order to investigate BBH mergers from strong interactions in GCs, we utilize results
from star cluster models that were evolved using the \texttt{MOCCA} (MOnte Carlo Cluster simulAtor) code
\citep[see][and reference therein for details about the \texttt{MOCCA} code and the Monte Carlo method]{Hypki2013,Giersz2013} as part of the MOCCA-Survey Database I project
comprising of nearly $2000$ GCs \citep{2017MNRAS.464L..36A}. \texttt{MOCCA} uses the orbit
averaged MC method \citep{Henon1971,Stodolkiewicz1986} to carry
out the long term evolution of spherically symmetric star clusters. For binary and stellar evolution, \texttt{MOCCA}
employs prescriptions provided by the \texttt{SSE/BSE} codes \citep{Hurley2000,Hurley2002}. In order to properly compute
strong binary-single and binary-binary interactions, \texttt{MOCCA} uses the \texttt{fewbody} code \citep{Fregeau2004} which is a
direct \textit{N}-body integrator for small \textit{N} systems. The MC method is significantly faster
than direct \textit{N}-body codes and \texttt{MOCCA} can simulate the
evolution of realistic GCs in a few days\footnote{To simulate a star cluster with a million objects using \texttt{MOCCA} on a present day single
CPU, single core processor, it is needed about a day, up to a week, depending on the initial conditions.}.
Comparisons between \texttt{MOCCA} and direct \textit{N}-body results show good agreement for both global parameters
and evolution of specific objects in GC models \citep{Giersz2013,Wang2016,2017MNRAS.470.1729M}.

\texttt{MOCCA} provides as an output every binary-single and binary-binary interaction that was computed
using the \texttt{fewbody} code. For this paper, we extracted all the strong binary-single interactions that take place within a Hubble time
and involve three BHs that individually have masses less than $100 M_{\odot}$. There were more than a million such interactions
from nearly $800$ models in the `MOCCA-Survey Database I'. Nearly all of these interactions ($99.8$\%) came from models in which
BH kicks were computed according to mass fallback prescription given by \citet{Belczynski2002}.

In the output data provided by \texttt{MOCCA}, all the parameters that were used to call the \texttt{fewbody} code for a particular
interaction are provided, including the impact parameter, relative velocity, BH masses and initial binary semi-major axis (SMA).
For the purpose of this study, we used the input parameters provided to \texttt{fewbody} for
a subsample of these million interactions to re-simulate these strong interactions with our 2.5 PN
few-body code described in \cite{2017ApJ...846...36S}, as further explained in Section \ref{sec:Re-simulations with a PN few-body code}.

\section{Numerical and Analytical Methods}\label{sec:Numerical and Analytical Methods}

In this section we describe our numerical and analytical methods used for estimating the
rate of eccentric BBH mergers forming through binary-single interactions extracted from `MOCCA-SURVEY Database I'.

\subsection{Re-Simulating with a PN Few-Body Code}\label{sec:Re-simulations with a PN few-body code}

All the binary-single interactions performed for `MOCCA-SURVEY Database I', were originally evolved
using the Newtonian few-body code {\texttt{fewbody}} \citep{Fregeau2004}.
To investigate the effects from GR, we re-simulated
these binary-single interactions with our 2.5 PN few-body code described in \cite{2017ApJ...846...36S}.
To this end, we first selected all the binary-single interactions from `MOCCA-SURVEY Database I' for which the
initial orbital energy is negative (GR effects are only important for hard-binary interactions \citep{2014ApJ...784...71S}),
and the tidal force exerted on the binary by the incoming single at peri-center assuming a Keplerian orbit
is larger than the binding force of the binary itself (dynamical BBH mergers only form through strong interactions).
This left us with a total of $\sim 500,000$ binary-single interactions.

For generating the initial conditions (ICs) for these interactions, we randomly sampled the respective phase angles according to
the orbital parameters \citep{Hut:1983js}, while keeping the initial binary SMA, eccentricity, impact parameter, and relative velocity fixed
to the values given by `MOCCA-SURVEY Database I'.
We did this $5$ times for each of the original $\sim 500,000$ binary-single interactions provided by \texttt{MOCCA} to achieve better statistics,
which then resulted in a total of $\sim 2.5\times10^{6}$ scatterings. Due to computational restrictions,
we had to limit each interaction to a maximum of $2500$ initial orbital times, which resulted in about $2\%$ unfinished interactions that we chose
to discard. Long duration interactions are usually a result of an interaction where
one of the three BHs is sent out on a nearly unbound orbit, and represents therefore not any special class of outcome \citep{2017arXiv170604672S}.
All results presented in this paper are based on the completed set of these interactions.

\begin{figure}
\centering
\includegraphics[width=\columnwidth]{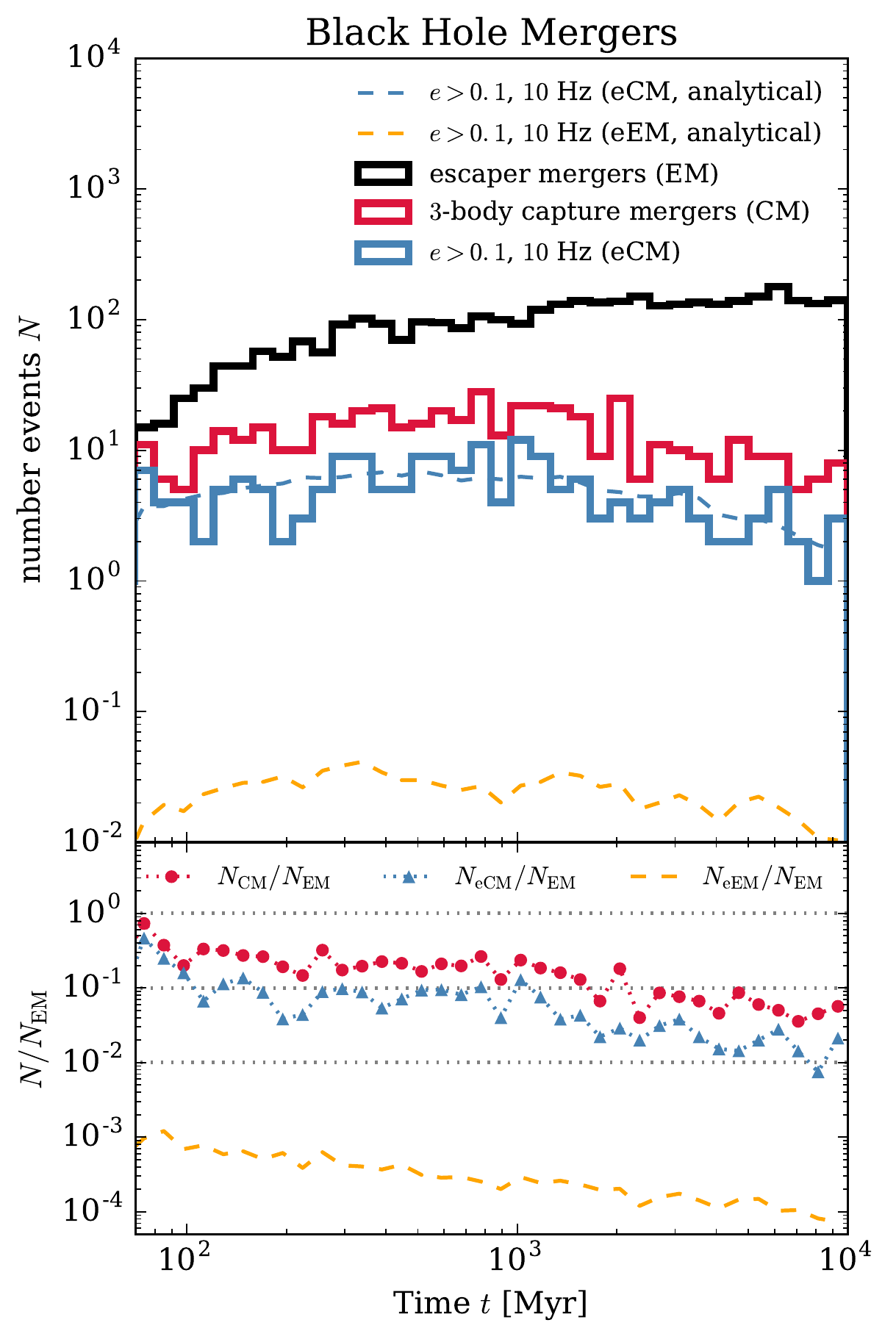}
\caption{Distribution of BBH mergers formed through binary-single interactions.
The results are based on $\sim 500,000$ binary-single interactions extracted from `MOCCA-SURVEY Database I',
each of which we simulated $5$ times using a 2.5 PN few-body code, as described in Section \ref{sec:Re-simulations with a PN few-body code}.
{\it Top plot}: Number of BBH mergers formed through binary-single interactions per logarithmic time interval as a function of time.
The {\it solid black} line shows the BBH mergers originating from the population kicked out of their host cluster through a binary-single
interaction. These BBHs are usually referred to as escapers.
The {\it dashed orange} line shows the BBH mergers with eccentricity $e>0.1$ at $10$ Hz derived from the escaper population (black solid line).
For this, we used our analytical framework described in Section \ref{sec:Analytical Estimations}.
The {\it solid red} line shows the BBH mergers formed through two-body GW captures in binary-single interactions, a population
we refer to as three-body GW capture mergers. Such mergers can only be probed using an $N$-body code that includes GW emission in the EOM.
The {\it solid blue} line shows the BBH mergers with eccentricity $e>0.1$ at $10$ Hz. As seen, this population is $\sim 100$ times larger than the
eccentric escaper population (dashed orange line), and is therefore completely dominated by three-body GW capture mergers.
The {\it dashed blue} line shows our analytical estimate of the three-body GW capture mergers with eccentricity $e>0.1$ at $10$ Hz,
as described in Section \ref{sec:Analytical Estimations}.
{\it Bottom plot}: Ratio between the outcomes from the top plot (dashed orange, solid red, solid blue) and the escaper population (solid black).
As seen, the three-body GW capture mergers constitute $\sim 10\%$ of all the BBH mergers observable at present time, where $1-5\%$ will have
an eccentricity $e>0.1$ at $10$ Hz. At early times the three-body GW capture mergers seem to even dominate the rate.}
\label{fig:BBHmergers}
\end{figure}

\subsection{Analytical Estimate}\label{sec:Analytical Estimations}

It has recently been illustrated that the distribution of eccentric BBH mergers forming through binary-single interactions
can be estimated analytically \citep{2014ApJ...784...71S, 2017arXiv170603776S, 2017arXiv171107452S}, despite the
highly chaotic nature of the three-body problem and the complexity of GR. In this section we describe how to apply these recent
calculations to estimate the population of eccentric mergers forming in GC data. For the equations
below we follow the notation from \cite{2017arXiv171107452S}, as well as assuming the equal mass limit.
This is an excellent approximation, as similar mass objects tend to interact at the same time
due to the effect of mass segregation \citep[e.g.][]{2016PhRvD..93h4029R}.

To estimate the number of GW capture mergers with measurable eccentricity $e_{f}$ at GW frequency $f$,
we first use that a typical binary-single BH interaction generally can be described as a series of temporary BBHs with a bound single BH \citep{2014ApJ...784...71S}.
The single and the BBHs exchange in a semi-chaotic way energy and angular momentum, which makes it possible
for the BBHs to occasionally reach very high eccentricities during the interaction \citep{2014ApJ...784...71S}. Now, if the eccentricity of a given temporary BBH
is high enough, the BBH will undergo a two-body GW capture merger while still being bound to the single: this is the population we loosely refer
to as {\it three-body GW capture mergers}. Although these mergers generally form
at very high eccentricity, they do not necessarily have a measurable eccentricity at the time of observation due to circularization during inspiral \citep{2017ApJ...840L..14S}.
Luckily, deriving the number of BBH mergers with $e_{f}$ at GW frequency $f$ is easier than deriving the full population of three-body GW capture
mergers \citep{2017arXiv171107452S}, which makes it possible to easily estimate their expected rate, as explained in the following.

Assuming $f$ only depends on peri-center distance \citep{Wen:2003bu, 2017arXiv171107452S}, a temporary BBH must form with a specific
peri-center distance $r_{\rm EM}$ (`EM' is short for `Eccentric Merger'), for its orbital eccentricity to be $e_{f}$ at frequency $f$.
The value for $r_{\rm EM}$ relates closely to the peri-center distance $r_{f}$ at which the GW frequency is $f$, a distance that
can be shown to fulfill $r_{f}^{3} \approx {2Gm}{f^{-2} {\pi}^{-2}}$, where $m$ is the mass of one of the three
(equal mass) BHs \citep[see e.g.][]{2017arXiv171107452S}. Using this approximation for
$r_{f}$, the relation between SMA and eccentricity derived by \cite{Peters:1964bc}, and that the initial orbital eccentricity
at $r_{\rm EM}$ is $\approx 1$ (a limit that follows from that $r_{\rm EM} \ll$ than the initial SMA), one now finds,
\begin{equation}
r_{\rm EM} \approx \left( \frac{2Gm}{f^2 {\pi}^2}\right)^{1/3} \frac{1}{2}  \frac{1+e_{f}}{e_{f}^{12/19}} \left[ \frac{425}{304} \left(1 + \frac{121}{304}e_{f}^2 \right)^{-1} \right]^{870/2299},
\label{eq:r_EM}
\end{equation}
as described in greater detail in \cite{2017arXiv171107452S}.
To clarify, our derived $r_{\rm EM}$ is the peri-center distance two BHs have to come within for their eccentricity to be $>e_{f}$ at frequency $f$.
Because $r_{\rm EM}$ is a fixed distance, the probability for a single temporary BBH to form with an initial peri-center distance $ < r_{\rm EM}$ is
simply $\approx {2r_{\rm EM}}/{a}$, where $a$ denotes the SMA of the initial target BBH, a relation that follows
from assuming the BBH eccentricity distribution is thermal \citep{Heggie:1975uy, 2017arXiv171107452S}.
Now, to find the probability for a single binary-single interaction to result in a BBH merger with an initial peri-center distance $<r_{\rm EM}$,
referred to as $P_{\rm EM}$, one simply needs to weight with the number of temporary
BBHs forming per binary-single interaction, a number we denote by $N_{\rm IMS}$, where `IMS' is short for `Intermediate State' \citep{2017arXiv171107452S}. From this
finally follows,
\begin{equation}
P_{\rm EM} \approx \frac{2r_{\rm EM}}{a} \times N_{\rm IMS},
\label{eq:Pef_a}
\end{equation}
where $N_{\rm IMS} \approx 20$ in the equal mass case \citep{2017arXiv170603776S, 2017arXiv171107452S}.
We have here assumed that if two BHs undergo an initial peri-center distance $ < r_{\rm EM}$ then they also merge, which
is an excellent approximation for sources observable by an instrument similar to the `Laser Interferometer Gravitational-Wave Observatory' (LIGO),
but not necessarily for sources in the frequency range of the `Laser Interferometer Space Antenna' (LISA).

We applied this analytical formalism to estimate the number of BBH mergers with eccentricity $>e_{f}$ at GW frequency $f$,
forming in the dataset `MOCCA-SURVEY Database I'. To this end,
we first calculated $P_{\rm EM}$ for each of the binary-single interactions in the set we extracted for re-simulation
(see Section \ref{sec:Re-simulations with a PN few-body code}), assuming that the three interacting BHs all
have the same mass equal to their average mass. As $P_{\rm EM}$ effectively describes the average number of
BBH mergers with eccentricity $>e_{f}$ at GW frequency $f$ forming per interaction, the distribution
of such mergers is simply given by the distribution of $P_{\rm EM}$. This approach
allows us to instantly derive simple relations between observed eccentricity and GW frequency, that otherwise would take
thousand of `CPU hours' and an extensive amount of coding. As shown in the sections below, the estimate from this analytical approach is remarkably
accurate.

\section{Results}\label{sec:Results}

Our main results are presented in Figure \ref{fig:BBHmergers}, where each of the shown outcomes are described in the paragraphs below.

\subsection{Escaping Black Hole Mergers}

The distribution of BBH mergers originating from the population of BBHs dynamically ejected from their host
cluster through binary-single interactions is shown in {\it black}. For this estimation, we first identified all the binary-single
interactions that resulted in an ejected BBH with a dynamical kick velocity (derived from the output of our re-simulated few-body interactions)
larger than the escape velocity of the cluster (derived from the central potential provided by the MOCCA-code output).
We then followed this escaped population using the orbital evolution equations given by \cite{Peters:1964bc},
from which we derived the final distribution of merger times. This population of BBH mergers have been extensively studied
using both $N$-body \citep[e.g.][]{2014MNRAS.440.2714B, 2017MNRAS.469.4665P} and MC \citep{2016PhRvD..93h4029R,2017MNRAS.464L..36A} techniques,
which all find that this dominates the present day BBH merger rate originating from GCs.

The distribution of BBH mergers with eccentricity $e>0.1$ at $10$ Hz originating from the escaper population (the one shown in black),
is shown with an {\it orange dashed} line. It was extremely difficult to numerically resolve this population due to its low statistics, so instead we
used our analytical framework described in Section \ref{sec:Analytical Estimations}.
To this end, we first selected all the binary-single interactions leading to an escaping BBH, after which we calculated the probability for each of these
to have $ < r_{\rm EM}$ using Equation \eqref{eq:Pef_a} with $N_{\rm IMS}$ set to $1$, as there is only one ejected BBH per interaction.
As seen, the fraction of escaping BBH mergers with $e>0.1$ at $10$ Hz is extremely low, which have
led several cluster studies to conclude that the rate of eccentric BBH mergers forming in GCs is far too low to be observable; however, as
described by \cite{2017ApJ...840L..14S, 2017arXiv171107452S}, the rate of eccentric mergers is not dominated by the escape merger
population, but instead by three-body GW capture mergers -- a statement recently confirmed by \cite{2017arXiv171204937R},
and further described below.

\subsection{Three-Body GW Capture Mergers}

The distribution of three-body GW capture mergers is shown in the top panel of Figure \ref{fig:BBHmergers} with a {\it red solid line}.
As seen, these three-body GW capture mergers constitute about $10\%$ of all the BBH mergers observable
at late times, and seem to even dominate the merger rate at early times. These results are in surprisingly good agreement with
recent analytical work by \cite{2017arXiv170603776S, 2017arXiv171107452S}.
The reason why the number of three-body GW capture mergers is surprisingly large, is because all binary-single interactions can contribute to
this merger population, and not only the ones leading to BBH escapers: from the data `MOCCA-SURVEY Database I'
we found that for every binary-single interaction leading to an escaper, there are of order $10^{2}$ binary-single interactions each of which
potentially can undergo a three-body GW capture merger without leading to an escaper.

The distribution of three-body GW capture mergers with eccentricity $e>0.1$ at $10$ Hz derived using our 2.5 PN few-body code,
is shown with a {\it blue solid line}. As seen, this population is much larger than the one originating from
the escaper population (orange dashed line), which clearly illustrates that the rate of eccentric sources is dominated
by three-body GW captures (to exactly which degree binary-binary interactions contribute is topic of current research). By comparing the blue and orange
histograms, one finds that the rate of eccentric sources increases by a factor of $\sim 100$ when three-body GW captures are included, which agrees
surprisingly well with the recent analytical derivations by \cite{2017arXiv171107452S}.
In short, this enhancement factor is a product of $N_{\rm IMS}$ (the three-body system has $N_{\rm IMS} \approx 20$ tries during the interaction per single escaper)
and a factor that represents the possibility for an eccentric three-body GW capture merger to form in binary-singles that do not lead
to an escaper (which is $\approx 5$).

The distribution of three-body GW capture mergers with eccentricity $e>0.1$ at $10$ Hz derived using our analytical framework
from Section \ref{sec:Analytical Estimations} is shown with a {\it dashed blue line}. As seen, the agreement with
our full numerical estimate (solid blue line) is remarkable, which proves the analytical framework as a highly useful tool
for exploring observable relations between eccentricity and GW frequency.

The bottom panel in Figure \ref{fig:BBHmergers} shows the different outcome distributions from the top plot divided by the distribution
of BBH mergers formed through escapers (orange/red/blue histograms divided by the black histogram).
As seen, the relative rate of eccentric (dominated by three-body GW captures) to circular mergers (dominated by the escapers)
is at present time $1-5\%$, which again is in excellent agreement with the analytical predictions by \cite{2017arXiv171107452S}. This leads to
the conclusion that the eccentric fraction is likely to be within observable limits if BBH mergers from GCs contribute notably to the observed population,
which highly motives further work on eccentric GW templates \citep[e.g.][]{2016PhRvD..94b4012H, 2016arXiv160905933H, 2017arXiv171106276H}.
Similar encouraging results are discussed in \cite{2017arXiv171204937R}.

\section{Conclusions}\label{sec:Conclusion}

We have in this paper presented estimates of the population of GW capture mergers forming during binary-single interactions (three-body GW capture mergers)
in GCs evolved using realistic prescriptions. To this end, we re-simulated $\sim 500,000$ strong binary-single interactions extracted from
the dataset `MOCCA-SURVEY Database I' derived using the MC code \texttt{MOCCA}, with a few-body code that includes
GW emission in the EOM \citep{2017ApJ...846...36S} using the PN formalism \citep[e.g.][]{2014LRR....17....2B}.
In addition, we further showed how the analytical framework from \cite{2017arXiv171107452S} can be used to make accurate and instant estimates
of the rate of BBH mergers that will appear in the observable GW band with a notable eccentricity.
This illustration provides an important piece in further developments of analytical GR models
for understanding the evolution of dense stellar systems.

Our analytical and numerical results strongly indicate that $\sim 10\%$ of all GC BBH mergers that are observable at present time
originate from three-body GW capture mergers (See bottom plot in Figure \ref{fig:BBHmergers}), which is in
excellent agreement with the recent analytical study by \cite{2017arXiv171107452S}; a result also confirmed by \cite{2017arXiv171204937R}.
In addition, the population of GC BBH mergers with eccentricity $>0.1$ at $10$ Hz is about $1-5\%$ of the total GC merger rate at present time,
which strongly suggests that eccentric mergers are within observable limits for an instrument similar to LIGO, given that GCs contribute to the
observed rate. This finding opens up for the possibility of using the eccentricity distribution to constrain the fraction of BBH mergers that form dynamically.
These promising results indeed motivate further work on the role of GR in the evolution of GCs \citep[e.g.][]{2006MNRAS.371L..45K, 2013MNRAS.434.2999B},
both from the numerical and the analytical sides.

\acknowledgments{Support for this work was provided by NASA through Einstein Postdoctoral
Fellowship grant number PF4-150127 awarded by the Chandra X-ray Center, which is operated by the
Smithsonian Astrophysical Observatory for NASA under contract NAS8-03060.
JS thanks the Niels Bohr Institute, the Kavli Foundation and the DNRF for
supporting the 2017 Kavli Summer Program, and the Nicolaus Copernicus Astronomical Center, Polish Academy of Sciences. JS and AA also thank the Flatiron Institute's Center for Computational Astrophysics for their generous support during the CCA Numerical Scattering Workshop.
AA and MG were partially supported by the Polish National Science Center (NCN), Poland, through the grant UMO-
2016/23/B/ST9/02732. AA is also supported by NCN through the grant UMO-2015/17/N/ST9/02573.}

%%%%%%%%%%%%%%%%%%%%%%%%%%%%%%%%%%%%%%%%%%%%%%%%%%%%%%%%%%%%%%%%%%%%%%%%   

%%%%%%%%%%%%%%%%%%%%%%%%%%%%%%%%%%%%%%%%%%%%%%%%%%%%%%%%%%%%%%%%%%%%%%%%   

\end{document}